# Extremal transmission and beating effect of acoustic wave in two-dimensional sonic crystal

Xiangdong Zhang[1] and Zhengyou Liu [2]

1*Deparment of Physics, Beijing Normal University, Beijing 100875, China;*
2*Key Lab of Acoustic and Photonic Materials and Devices of Ministry of Education, Wuhan University, Wuhan 430072, China and Department of Physics, Wuhan University, Wuhan 430072, China*

## Abstract

The extremal transmission of acoustic wave near the Dirac point in two-dimensional (2D) sonic crystal (SC), being inversely proportional to the thickness of sample, has been demonstrated experimentally for the first time. Some unusual beating effects have been observed experimentally when the acoustic pulse transport through the 2D SC slabs. Such phenomena are completely different from the oscillations of the wave in a slab or cavity originating from the interface reflection or Fabry-Perot effect. They can be regarded as acoustic analogue to Zitterbewegung of relativistic electron. The physical origination for the phenomenon has been analyzed.

**PACS numbers:**   43.35.+d, 63.20.–e

Recently, there has been a great deal of interest in studying the transport of electromagnetic wave near the Dirac point in two-dimensional (2D) photonic crystal (PC) [1-3]. In some 2D PCs with triangular or honeycomb lattices, the band gap may become vanishingly small at corners of the Brillouin zone, where two bands touch as a pair of cones. Such a conical singularity is referred to as the Dirac point similar to the case of electron in the monolayer graphene [4, 5]. When we describe the transport of electromagnetic wave near the Dirac point inside the PC, the Maxwell equations can reduce to the Dirac equation due to the linear dispersion around it [1]. This leads to some unusual transmission properties of electromagnetic wave in the PC [1-3]. For example, the optical conductance near the Dirac point is inversely proportional to the thickness of sample, which is different from the ballistic behavior in general band regions and exponential decay with the increase of sample thickness when the frequency lies in the gap regions [2]. Although such an extremal transmission phenomenon in the 2D PC has been pointed out theoretically [2], it has not been observed experimentally so far.

In this work, we demonstrate both experimentally and theoretically that such a phenomenon can also appear for acoustic wave in 2D sonic crystal (SC). Based on these, we study experimentally dynamic behavior of pulse transmission through the 2D SC slabs with the center frequency near the Dirac point. Some unusual beating effects are found. The phenomenon can be regarded as acoustic analogue to Zitterbewegung (ZB) of relativistic electron, which is similar to the case of photon near the Dirac point in the PC [3].

The SC used in the experiments consists of a hexagonal array of steel cylinders immersed in water with a lattice constant a=1.5mm and a filling ratio 0.403. The band structure of the system obtained by the multiple-scattering Korringa–Kohn–Rostoker method [6] is shown in Fig. 1 (a). The material parameters used in the calculations are: density $\rho = 1000 g/m^3$ and sound velocity $c = 1490 m/s$ for water, and density $\rho = 7800 g/m^3$ and longitudinal wave velocity $c_l = 6010 m/s$ for steel. The corresponding measured data of transmission coefficient along $\Gamma K$ direction for a rectangular sample with different thickness and width 86 mm are plotted in Fig. 1(b). The measurement, based on the well-known ultrasonic transmission technique [7], is performed on a Panametrics LSC-02 ultrasonic scanning system. A pulser/receiver generator (Panametrics model 5900PR) produces a driving short-duration pulse of a width about 2.0 μs for the generating transducer. The generating transducer, placed far away from the sample to yield an input pulse approximating a plane wave, serves as the source of acoustic waves, and a receiving transducer placed on the opposite side of the sample serves as the detector for the transmitted signal. Both the generating and receiving transducers have a diameter of 25mm and a central frequency of 0.5 MHz. The assembly of the generating transducer, the sample and the receiving transducer is immersed in a water tank. The excellent agreement between the experimental measurements and the band structure can be observed.

The key feature of this band structure is that the band gap becomes vanishingly small at corners of the Brillouin zone at $f = 0.55 MHz$, where two bands touch as a pair of cones in a linear fashion. This is similar to the case of the Dirac point of photon in the PC although some unsymmetries exist [1, 2]. If we regard such a point as the Dirac point of acoustic wave in the SC, the transport of acoustic wave around it can be described by the following Dirac equation [1, 2]

$$H \begin{pmatrix} \psi_1 \\ \psi_2 \end{pmatrix} = \delta\omega \begin{pmatrix} \psi_1 \\ \psi_2 \end{pmatrix}, \delta\omega = \omega - \omega_D, \qquad (1)$$

with

$$H = \begin{pmatrix} 0 & -iv_D(\partial_x - i\partial_y) \\ -iv_D(\partial_x + i\partial_y) & 0 \end{pmatrix}. \qquad (2)$$

Where $\psi = (\psi_1, \psi_2)$ represents the amplitudes of two degenerate Bloch states at one of the corners of the hexagonal first Brillouin zone. The frequency $\omega_D$ and velocity $v_D$ in the Dirac point depend on the parameters of the SC. Thus, the flux of acoustic wave inside the SC can be written as $j_D = v_D(\psi_1^*\psi_2 + \psi_2^*\psi_1)$. In the free space outside of the SC, the transport of acoustic wave still follows the elastic wave equation. If the acoustic flux outside of the SC is defined as $j_P$, applying the boundary condition of flux conservation ($j_P = j_D$) for the SC slabs with thickness L, we can obtain the transmission (T) of the acoustic wave around the Dirac point $T \propto 1/L$. The detailed derivation is similar to the case of electromagnetic wave in Ref.[2]. This means that the transmission of acoustic wave near the Dirac point is inversely proportional to the thickness of sample, which is similar to diffusion behavior of wave through a disordered medium even in the absence of any disorder in the SC.

In order to test such a phenomenon experimentally, we fabricate a series of SC slabs of different thicknesses and measure the transmission of acoustic wave through them. The widths of all samples are taken as 86mm. The product of transmission coefficient along $\Gamma K$ direction and $L$ as a function of the sample thickness at $f = 0.55 MHz$ is plotted in Fig.2 as dark dots. The red solid line represents the numerical results obtained by a rigorous multiple-scattering method [6]. Excellent agreement between the measured and the calculated results is apparent. Both of them show that $T^2 L$ keeps constant (green line) with the increase of $L$ although it oscillates around green line. Here the green line is drawn only for view, which can be regarded as the corresponding to the above theoretical result ($T \propto 1/L$). The oscillating characteristics of $T^2 L$ depend on the interface and the thickness of the sample, but they leave the green line unaffected [2]. That is to say, the transmission of acoustic wave in such a case is actually inversely proportional to the thickness of the sample. This is completely different from the ballistic behavior in general band regions, where the transmission coefficient is near constant and $T^2 L$ increase linearly with the increase of $L$. At the same time, it is also different from the exponential decay with the increase of sample thickness when the frequency lies in the gap regions. This means that we have demonstrated experimentally the existence of extremal transmission of acoustic wave near the Dirac point in the SC. Such an extremal transmission can appear only in small frequency interval around the Dirac point, which is similar to the case of electromagnetic wave in the PC [2]. It is suitable for the transmission along $\Gamma K$ direction. For the transmission along $\Gamma M$ direction, it is suppressed exponentially with increasing the sample thickness because there exists a gap at the same frequency.

After the phenomenon of extremal transmission has been demonstrated, in the following we investigate the dynamic behavior of acoustic wave transport near the Dirac point in the above SC. Thus, we inject a Gaussian acoustic pulse into the samples and measure the time dependence of transmission through the sample. The measured results for filter frequency widths $\Delta f = 0.01 MHz$, $0.02 MHz$ and $0.03 MHz$ are plotted in Fig.3 (a), (b) and (c). Dark, red and green correspond to the results with L=14a, 19a and 24a, respectively. It is seen clearly that all of them are oscillations (beating effect). The oscillations have a transient character, as they are attenuated exponentially with the increase of time. They can be observed only when the thickness of the sample is bigger than several wavelengths such as 10a. For thin samples such as $L \leq 3a$ (the sample thickness near one wavelength), the second peak basically disappears. That is to say, no oscillation can be observed. However, once the oscillation appears, its period depends weakly on the thickness of the sample, while it is linearly dependent on the filter frequency widths. For example, the period for

$\Delta f = 0.01 MHz$ (in Fig.3 (a)) is nearly 3 times of that for $\Delta f = 0.03 MHz$ (Fig.3 (c)).

It is well known that the oscillations of the transmission intensity can also appear when the waves transport through a slab or cavity due to the interface reflection or Fabry-Perot effect [8, 9]. In this case, the oscillation period is determined by the thickness of the sample and the central frequency of the pulse. The frequency need not locate at the vicinity of the Dirac point. When the frequency (or wavelength) and the thickness of the sample or cavity satisfy the resonant conditions, the oscillation appears [8, 9]. However, the present oscillations are in contrast to such a case. This means that such an oscillation does not originate from the interface or Fabry-Perot effect, although they maybe have some effect on such a phenomenon. It is also not related to the localization effect, because the transport of wave exhibits diffusion behavior. Such a phenomenon is directly related to the ZB, which is similar to the case of electromagnetic wave [3].

ZB represents an oscillatory motion of free electrons described by the Dirac equation in the absence of external fields, which is caused by the interference between the positive and negative energy states [9]. It was believed that the experimental observation of the effect is impossible since one would confine the electron to a scale of the Compton wavelength [11]. Although many theoretical investigations have shown that such a phenomenon can be also found for nonrelativistic electrons in solid systems [12-15] and photon in the PC [3], it has not been observed directly from experimental measurements all the time.

According to the theory of ZB, the evolution of wave packet with the time in the systems can be obtained analytically. For example, we consider a Gaussian wave packet in the form [14, 18, 3]

$$\psi(\vec{r},0) = \frac{1}{2\pi} \frac{d}{\sqrt{\pi}} \int d^2\vec{k} e^{-\frac{1}{2}d^2 k_x^2 - \frac{1}{2}d^2(k_y - k_{y0})^2} e^{i\vec{k}\cdot\vec{r}} \begin{pmatrix} 1 \\ 0 \end{pmatrix}. \qquad (3)$$

The packet is centered at $k_0 = (0, k_{0y})$ and is characterized by a width of d. The unit vector (1, 0) is a convenient choice [18, 3]. Here $\vec{r}$ and $\vec{k}$ represent the position and wave vector of the wave packet, respectively, $k_x$ and $k_y$ are two components of the wave vector. By using the Hamiltonian in Eq.(2), the time-dependent displacement ($\overline{x}(t)$) of such a Gaussian wave packet along $\Gamma K$ direction of the SC can be obtained as [13,17, 3]

$$\overline{x}(t) = \frac{d^2}{\pi} \int_{-\infty}^{\infty} \int_{-\infty}^{\infty} \frac{k_y}{2k^2}[1 - \cos(2v_D kt))] e^{-d^2 k_x^2 - d^2(k_y - k_{y0})^2} dk_x dk_y. \qquad (4)$$

After the average displacement is obtained, the velocity of wave packet can be calculated by $\bar{v}(t) = \partial \bar{x}(t)/\partial t$. The calculated result for $\bar{v}(t)$ as a function of time is plotted in Fig.4 as solid line. Here $v_D$ is obtained experimentally by calculating the ratio between transport distance of acoustic pulse and time, $k_{0y}$ and d are taken as corresponding to $\Delta f = 0.02 MHz$. It can be seen clearly that it is oscillation function of time, which is a direct manifestation of ZB. The period of ZB is linearly dependent on $k_{0y}$ or $\delta\omega = 2\pi\Delta f$ due to linear Dirac-like dispersion around the Dirac point. The amplitude of ZB has a transient character, as it is attenuated exponentially with the increase of time. This can be understood from Eq.(4), in which there are terms for exponential decay. Although these oscillation features for $\bar{v}(t)$ are agreement qualitatively with the experimental results for the transmission coefficient, it seems to be not interrelated between them. In fact, they are corresponding. This is because the velocity of wave packet $\bar{v}(t)$ is proportional to the energy flux ($j(t)$) of acoustic pulse ($j(t) \propto \bar{v}(t)$). At the same time, $j(t)$ can be obtained by making a difference of pulse intensity (I(t)) to time ($j(t) = dI(t)/dt$). The intensity (I(t)) or transmission coefficient can be obtained from experimental measurements (see Fig.3). Thus, we can obtain $j(t)$ from the experimental data and quantitatively compare it with the theoretical result for $\bar{v}(t)$.

Dark dots (circle and triangular) in Fig.4 are the results by making a difference of the experimental data of transmission coefficient to time. That is to say, they are the experimental results. Here the data is only taken for one period (between the second peak and the third peak in Fig.3 (b)). Dark triangular and circle correspond to the case with L=14a and 24a, respectively. Comparing them with solid line, we find that the oscillating periods are basically identical. Some differences between the theoretical results and experimental measurements originate from the effect of finite size and interface as has been analyzed above. With the increase of sample thickness, the experimental results become more agreement with the theoretical estimation.

In addition, the deficiency of the oscillation for the thin samples can be also explained by the theory of ZB. When the thickness of the sample is very small, the transport of wave exhibits ballistic behavior. In this case, it can not be described by Eq.(1). Therefore, the oscillation disappears. When the thickness of the sample becomes bigger, diffusion behavior plays a leading role. In such a case, remarkable oscillations can be observed as shown in Fig.3. In a word, all phenomena observed in the experiments can be explained by the theory of ZB. This means that the beating effect observed from the experimental measurements is acoustic analogue to the ZB of relativistic electron, which is similar to the case of electromagnetic wave in the PC [3] or electron in the monolayer graphene [16-17].

The above experiments and calculations only focus on one kind of structure parameters of the SC. In fact, we have also considered the other cases with different structure parameters, for example, the radius of steel cylinder change from 0.3a to 0.4a. The similar phenomena have been obtained. This indicates that some phenomena in relativistic quantum mechanics, which is very difficult to be observed experimentally in the electron systems, can be observed easily by experimental measuring through acoustic wave in the SC slabs.

In summary, we have demonstrated experimentally existence of a new transport regime of acoustic wave in the SC near the Dirac point. In this regime, the transmission of acoustic wave is inversely proportional to the thickness of sample, which can be described by the Dirac equation very well. Based on these, we have observed directly acoustic analogue to the ZB of relativistic electron from experimental measurements for the time dependence of transmission coefficients of acoustic pulse. We anticipate our work to be a starting point for more experimental investigations on the phenomena of relativistic quantum mechanics by using classical waves and extensive applications of the phenomena to acoustic devices.

## Acknowledgements

This work was supported by the National Natural Science Foundation of China (Grant Nos. 10674017 and 50425206) and the National Key Basic Research Special Foundation of China under Grant 2007CB613205. The project was also sponsored by NCET and RFDP.

## References

[1] F. D. M. Haldane and S. Raghu, Phys. Rev. Lett. **100,** 013904 (2008); S. Raghu and F. D. M. Haldane, e-print arXiv:condmat/0602501.

[2] R. A. Sepkhanov, Ya. B. Bazaliy, and C. W. J. Beenakker, Phys. Rev. A **75**, 063813 (2007); X.D. Zhang, Phys. Lett. A (to be published).

[3] X.D. Zhang, Phys. Rev. Lett. **100,** 113903 (2008).

[4] K. S. Novoselov, A. K. Geim, S. V. Morozov, D. Jiang, Y. Zhang, S. V. Dubonos, I. V. Grigorieva,

and A. A. Firsov, Science 306, 666 (2004).

[5] T. Ando, J. Phys. Soc. Jpn. 74, 777 (2005).

[6] J. Mei, Z. Liu, J. Shi and D. Tian, Phys. Rev. B 67, 245107 (2003); Y. Lai, X. Zhang, and Z. Q. Zhang, Appl. Phys. Lett. 79, 3224 (2001).

[7] M. Ke, Z. He, S. Peng, Z. Liu, J. Shi, W. Wen and P. Sheng, Phys. Rev. Lett. 99, 044301 (2007).

[8] Helios Sanchis-Alepuz, Yuriy A. Kosevich, and Jose Sanchez-Dehesa, Phys. Rev. Lett. **98**, 134301 (2007).

[9] Z. He, S. Peng, F. Cai, M. Ke, and Z. Liu, Phys. Rev. E, **76**, 056605 (2007).

[10] E. Schrödinger, Sitzungsber. Preuss. Akad. Wiss., Phys. Math. Kl. 24, 418 (1930).

[11] H. Feschbach and F. Villars, Rev. Mod. Phys. 30, 24 (1958); A. O. Barut and A. J. Bracken, Phys. Rev. D 23, 2454 (1981); A. O. Barut and W. Thacker, Phys. Rev. D 31, 1386 (1985); K. Huang, Am. J. Phys. 20, 479 (1952).

[12] F. Cannata, L. Ferrari, and G. Russo, Solid State Commun. 74, 309 (1990); L. Ferrari and G. Russo, Phys. Rev. B 42, 7454 (1990); F. Cannata and L. Ferrari, Phys. Rev. B 44, 8599 (1991).

[13] W. Zawadzki, Phys. Rev. B 72, 085217 (2005); T. M. Rusin and W. Zawadzki, J. Phys.: C19, 136219 (2007)

[14] J. Schliemann, D. Loss, and R. M. Westervelt, Phys. Rev. Lett. 94, 206801 (2005); Phys. Rev. B 73, 085323 (2006).

[15] S.-Q. Shen, Phys. Rev. Lett. 95, 187203 (2005).

[16] M. I. Katsnelson, Eur. Phys. J. B 51, 157 _2006_.

[17] J. Cserti and G. David, Phys. Rev. B 74, 172305 (2006).

[18] T. M. Rusin and W. Zawadzki, Phys. Rev. B **76**, 195439 (2007).

# Figure Captions

Fig. 1 (a) Calculated photonic band structure of acoustic wave for a triangular lattice of steel cylinder with R/a=1/3 in a water background. (b) The measured transmission coefficient of acoustic wave along $\Gamma K$ direction for 2D sonic crystal corresponding to (a). The solid line corresponds to the result with the sample thickness $L = 10a$ and dashed line to that with $L = 20a$.

Fig.2 (a) Schematic picture depicting measure processes. (b) The product of $T^2$ and $L$ as a function of sample thickness at $f = 0.55 MHz$. Dark dots correspond to experimental data and red line to numerical results. The green line is drawn only for view, which can be regarded as the corresponding to the theoretical result $T \propto 1/L$. The other parameters are identical to those in Fig.1.

Fig.3 Experimental data of the transmission after the injection of a Gaussian pulse with center frequency $f = 0.55 MHz$ into the samples. Dark, red and green correspond to the cases with L=14a, 19a and 24a, respectively. (a),(b) and (c) correspond to the cases with the filter frequency widths $\Delta f = 0.01 MHz$, $0.02 MHz$ and $0.03 MHz$, respectively.

Fig.4 Comparison between the theoretical result (solid line) for velocity ($\bar{v}(t)$) of pulse evolution and experimental measurements (circle, triangular dots) for energy flux $j(t)$. Triangular and circle dots correspond to the cases with L=14a and 24a for the filter frequency widths $\Delta f = 0.02 MHz$, respectively. The other parameters are identical to those in Fig.3.

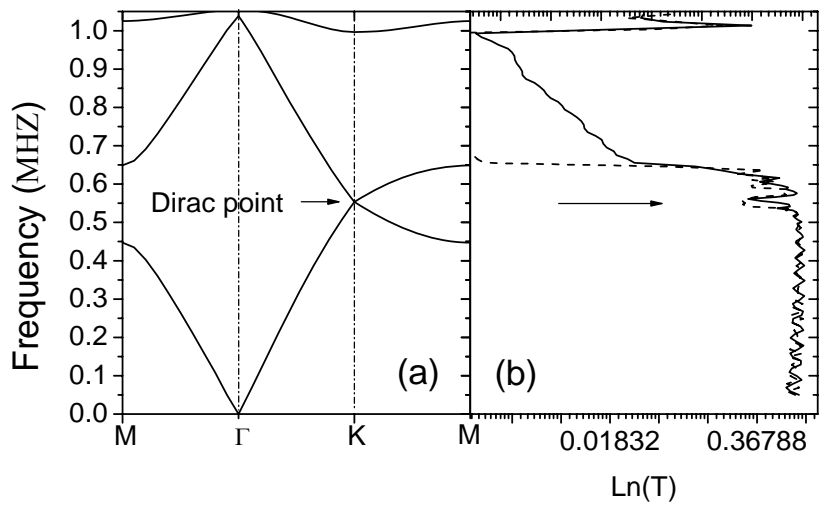

Fig.1

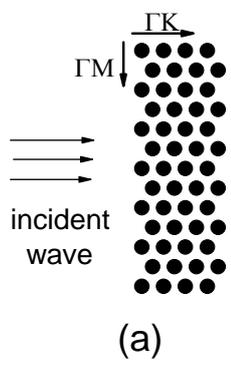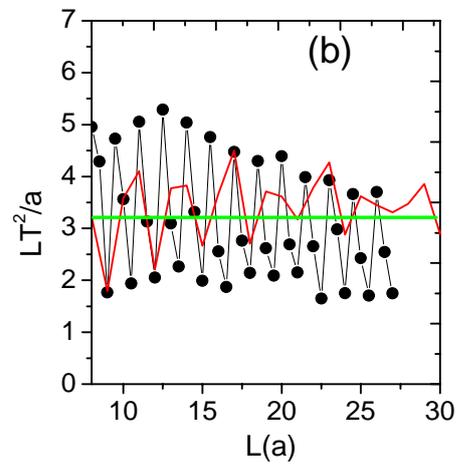

Fig.2

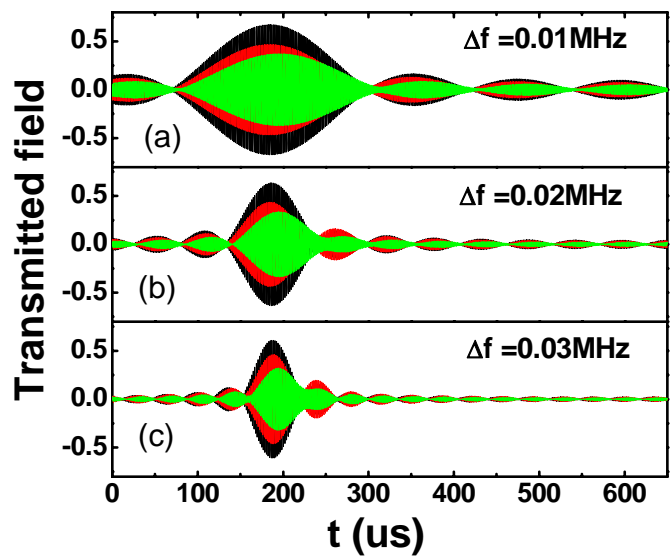

Fig.3

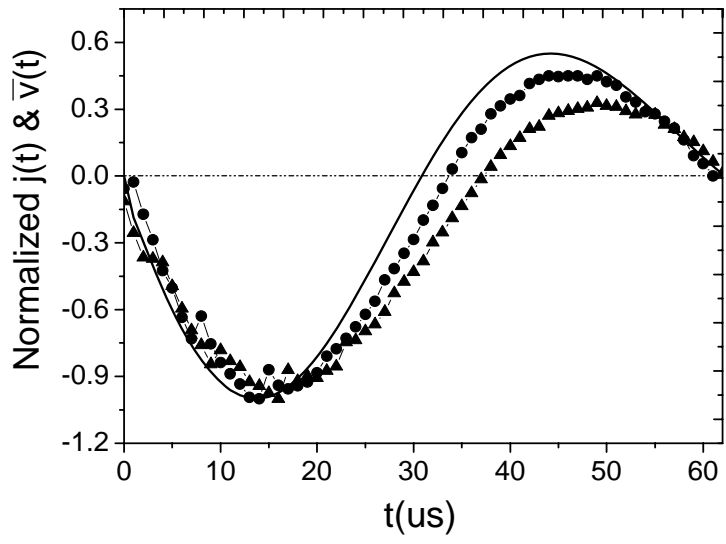

Fig.4